\documentclass{article}

\usepackage{amssymb}

\def\real{\mathbb{R}}
\def\lbag{\lbrack\!\lbrack}
\def\rbag{\rbrack\!\rbrack}

\def\mod{\mathit{mod}}
\def\sgn{\mathit{sgn}}
\def\ext{\mathit{vert}}
\def\ray{\mathit{ray}}
\def\dray{\mathit{line}}
\newcommand{\false}{\mathit{false}}
\newcommand{\true}{\mathit{true}}
\newcommand*{\Lin}[1][]{{\mathit{Lin}_{#1}}}
\newcommand{\con}{\mathit{connect}}
\newcommand{\conv}{\mathit{conv}}
\newcommand{\cone}{\mathit{cone}}
\newcommand{\extreme}{\mathit{extreme}}
\newcommand{\hull}{\mathit{hull}}
\newcommand{\res}{{E_\mathrm{res}}}
\newcommand{\preRay}{d_{\mathit{pre}}}
\newcommand{\postRay}{d_{\mathit{post}}}
\newcommand{\intersect}{\mathit{intersect}}

\newcommand{\ang}[2]{{#1 \measuredangle #2}}
\newcommand*{\eplus}[1][1]{{e_{(i+#1) \; \mod \; n}}}

\newcommand{\coincide}{\mathrm{c}}
\newcommand{\seq}[1]{{\langle #1 \rangle}}
\newcommand{\inBox}{\mathit{inBox}}
\newcommand{\add}{\mathit{add}}
\newcommand{\saturates}{\mathit{saturates}}
\newcounter{lineNumber}
\newcommand{\lineno}{\stepcounter{lineNumber}\arabic{lineNumber} }

\newlength{\ii}
\newtheorem{theorem}{Theorem}[section]
\newtheorem{lemma}{Lemma}[section]

\newtheorem{proposition}{Proposition}[section]

\newenvironment{proof}{{\noindent \bf Proof.}}{$\blacksquare$\vspace{0.5cm}}

\begin{document}

\title{Convex Hull of Planar H-Polyhedra}

\author{Axel Simon and Andy King\\
Computing Laboratory\\
University of Kent, CT2 7NF, UK\\
$\{$a.simon,a.m.king$\}$@ukc.ac.uk}

\maketitle

\begin{abstract}
  Suppose $\langle A_i, \vec{c}_i \rangle$ are planar (convex) H-polyhedra,
  that is, $A_i \in \real^{n_i \times 2}$ and $\vec{c}_i \in
  \real^{n_i}$.  Let $P_i = \{ \vec{x} \in \real^2 \mid A_i\vec{x}
  \leq \vec{c}_i \}$ and $n = n_1 + n_2$.  We present an $O(n \log n)$
  algorithm for calculating an H-polyhedron $\langle A, \vec{c}
  \rangle$ with the smallest $P = \{ \vec{x} \in \real^2 \mid A\vec{x}
  \leq \vec{c} \}$ such that $P_1 \cup P_2 \subseteq P$.

\end{abstract}


\noindent {\bf C.~R.~Categories:}  
I.3.5 [Computational Geometry and Object Modeling]: 
Boundary representations; 
Geometric algorithms, languages, and systems, 
I.3.6 [Methodology and Techniques]: 
Graphics data structures and data types,
F.3.1 [Specifying and Verifying and Reasoning about Programs]:
Invariants, 
Mechanical verification. 

\thispagestyle{empty}

\section{Introduction}

The convex hull problem is classically stated as the problem of
computing the minimal convex region that contains $n$ distinct points
$\{ \langle x_i, y_i \rangle \}_{i=1}^{n}$ in the Euclidean plane
$\real^2$.  The seminal work of Graham \cite{graham72efficient} showed
that the convex hull problem can be solved in $O(n \log n)$ worse-case
running time.  It inspired many to elaborate on, for example, the
three and more dimensional case, specialised algorithms for polygons,
on-line variants, {\it etc.}
\cite{preparata85computational,seidel97handbook}.  The convex hull of
polytopes (bounded polyhedra) can be calculated straightforwardly by
taking the convex hull of their extreme points. However, calculating
the convex hull for polyhedra turns out to be more subtle due to a
large number of geometric configurations.  Even for planar polyhedra,
the introduction of rays makes it necessary to handle polyhedra such
as a single half-space, a single ray, a single line, two facing (not
coinciding) half-spaces, etc., all of which require special handling
in a point-based algorithm. The problem is exacerbated by the number
of ways these special polyhedra can be combined.  In contrast, we
present a direct reduction of the convex hull problem of planar
polyhedra to the convex hull problem for a set of points
\cite{graham72efficient}.  By confining all input points to a box and
applying the rays to translate these points outside the box, a linear
pass around the convex hull of all these points is sufficient to
determine the resulting polyhedron.  By adopting the classic Graham
scan algorithm, our algorithm inherits its $O(n \log n)$ time
complexity.  The standard tactic for calculating the convex hull of
$H$-polyhedra is to convert the input into an intermediate ray and
vertex representation.  Two common approaches to this conversion
problem are the double description method \cite{motzkin53double} (also
known as the Chernikova algorithm \cite{chernikova68algorithm}) and
the vertex enumeration algorithm of Avis and Fukuda
\cite{avis92pivoting}.  The Chernikova method leads to a cubic time
solution for calculating the convex hull of planar $H$-polyhedra
\cite{verge92note} whereas the Avis and Fukuda approach runs in
quadratic time.

The remaining sections are organised as follows: A self-contained
overview of the algorithm, together with a worked example, is given in
the next Section.  A formal proof of correctness is given in Section
\ref{sec-correct}.  Section \ref{sec-conclusion} concludes.

\section{Planar Convex Hull Algorithm\label{sec-alg}}

The planar convex hull algorithm takes as input two $H$-polyhedra and
outputs the smallest $H$-polyhedron which includes the input.  The
$H$-representation of a planar polyhedron corresponds to a set of
inequalities each of which takes the form $ax + by \leq c$, where
$a,b, c \in \real$ and either $a \neq 0$ or $b \neq 0$.  Let $\Lin$
denote the set of all such inequalities.  The vector $\langle a,b
\rangle$ is orthogonal to the boundary of the halfspace induced by $a
x + b y \leq c$ and points away from the feasible space.  This vector
induces an ordering on halfspaces via the orientation mapping
$\theta$.  This map $\theta : Lin \to [0,2\pi)$ is defined such that
$\theta(ax+by\leq c) = \psi$ where $\cos(\psi) = a/\sqrt{a^2+b^2}$ and
$\sin(\psi) = b/\sqrt{a^2+b^2}$.  The mapping $\theta$ corresponds to
the counter-clockwise angle which the half-space of $x \leq 0$ has to
be turned through to coincide with that of $a x + b y \leq c$.
Sorting half-spaces by angle is the key to efficiency in our
algorithm.  However, $\theta$ is only used for comparing the
orientation of two half-spaces. To aid the explanation of the
algorithm, the concept of angular difference $\ang{e_1}{e_2}$ between
two inequalities $e_1$ and $e_2$ is introduced as the
counter-clockwise angle between $\theta(e_1)$ and $\theta(e_2)$. More
precisely $\ang{e_1}{e_2} = (\theta(e_2) - \theta(e_1)) \; \mod \;
2\pi$.  Note that this comparator can be realized without recourse to
trigonometric functions \cite{sedgewick98algorithms}.

\begin{figure}[t]
  \fbox{
    \begin{minipage}{\ii}
      \begin{tabbing}
        99 \=xxx\=xxx\=xxx\=xxx\=xxx\=xxx \kill
        \lineno \> function $\extreme(E)$ begin \\
        \lineno \> \> sort $E$ to obtain $e_0 \ldots, e_{n-1}$
        such that $\theta(e_0) < \theta(e_1) < \ldots <
        \theta(e_{n-1})$;\\
        \lineno \> \> $V$ := $R$ := $\emptyset$; \\
        \lineno \> \> if $E = \{ ax + by \leq c\}$ then $R$ := $\{
        \langle -a /\sqrt{a^2+b^2} , -b /\sqrt{a^2+b^2}  \rangle
        \}$;\\
        \lineno \> \> for $i \in [0, n-1]$ do let $e_{i} \equiv ax
        + by \leq c$ in begin \\
        \lineno \> \> \> // are the intersection points of this
        inequality degenerated?\\
        \lineno \> \> \> $\preRay$ := $(\theta(e_{i})-\theta(e_{i-1
          \; \mod \; n}))\; \mod \;2\pi \geq \pi \vee n=1$; \\
        \lineno \> \> \> $\postRay$ :=  $(\theta(e_{i+1 \; \mod \; n}) -
        \theta(e_{i})) \; \mod \; 2\pi \geq \pi \vee n=1$; \\
        \lineno \> \> \> if $\preRay$ then $R$ := $R \cup \{ \langle b
        /\sqrt{a^2+b^2}, -a /\sqrt{a^2+b^2} \rangle \}$;\\
        \lineno \> \> \> if $\postRay$ \= then $R$ := $R \cup \{
        \langle -b /\sqrt{a^2+b^2}, a  /\sqrt{a^2+b^2} \rangle \}$;\\
        \lineno \> \> \> \> else $V$ := $V \cup \intersect( e_{i},
        e_{(i+1) \; \mod \; n})$; \\
        \lineno \> \> \> if $\preRay \wedge \postRay$ then $V$ :=
        $V \cup \{v\}$      where $v \in \{ \langle x , y \rangle \mid
        a x + b y = c \}$\\
        \lineno \> \> end \\
        \lineno \> \> return $\langle V, R \rangle$ \\
        \lineno \>end\\
        \lineno \>\\
        \lineno \> function $\con(\langle x_1, y_1 \rangle,
        \langle x_2, y_2 \rangle)$\\
        \lineno \> \> return  $(y_2 - y_1) x + (x_1 - x_2) y \leq (y_2 -
        y_1) x_1 + (x_1 -x_2)y_1$\\
        \lineno \>\\
        \lineno \> function $\saturates(\langle x_1, y_1 \rangle,
        ax + by \leq c)$\\
        \lineno \> \> return $(a x_1 + b y_1 = c)$ \\
        \lineno \>\\
        \lineno \> function $\inBox(s, \langle x, y \rangle)$\\
        \lineno \> \> return  $|x| \! < \! s \wedge |y| \! <
        \! s$
      \end{tabbing}
    \end{minipage}}
    \caption{Convex hull algorithm for planar
      polyhedra}\label{fig-aux}
\end{figure}

The algorithm makes use of a number of simple auxiliary functions.
The function
$\intersect(a_1 x + b_1 y \leq c_1,\; a_2 x + b_2 y \leq c_2)$
calculates the set of intersection points of the two lines $a_1 x +
b_1 y = c_1$ and $a_2 x + b_2 y = c_2$. In practice an implementation
of this function only needs to be partial since it is only applied in
the algorithm when the result set contains a single point.  
The remaining auxiliaries are listed in Figure~\ref{fig-aux}.
The $\con$ function generates an inequality from two points subject to the
following constraints: the halfspace induced by $\con(p_1, p_2)$ has
$p_1$ and $p_2$ on its boundary and if $p_1, p_2, p_3$ are ordered
counter-clockwise then $p_3$ is in the feasible space.  The notation
$\overline{p_1, p_2}$ is used to abbreviate $\con(p_1, p_2)$.
Furthermore, the predicate $\saturates(p,e)$ holds whenever the point
$p$ is on the boundary of the halfspace defined by the inequality $e$.
Finally, the predicate $\inBox(s, p)$ determines whether the point $p$
occurs within a square of width $2s$ that is centred on the origin.

\begin{figure}
  \fbox{
    \begin{minipage}{\ii}
      \begin{tabbing}
        99 \=xxx\=xxx\=xxx\=xxx\=xxx\=xxx \kill
        \lineno \> function $\hull(E_1, E_2)$ begin \\
        \lineno \> \> // assertion: each $E_i$ is satisfiable and
        non-redundant\\
        \lineno \> \> if $E_1 = \emptyset \vee E_2 = \emptyset$
        then return $\emptyset$; \\
        \lineno \> \> $\langle P_1, R_1 \rangle$ :=
        $\extreme(E_1)$; \\
        \lineno \> \> $\langle P_2, R_2 \rangle$ :=
        $\extreme(E_2)$;\\
        \lineno \> \> $P$ := $P_1 \cup P_2$; \\
        \lineno \> \> $R$ := $R_1
        \cup R_2$; // Note: $|R| \leq 8$\\
        \lineno \> \> $s$ := $\max\{ |x|, |y| \mid \langle x, y
        \rangle \in P \}$ + 1; \\
        \lineno \> \> // add a point along the ray, that goes through
        $x,y$\\
        \lineno \> \> // and is outside the box\\
        \lineno \> \> $Q$ := $P$;\\
        \lineno \> \> for $\langle x, y, a, b \rangle \in P \times R$ do
        $Q$ := $Q \cup \{ \langle x + 2\sqrt{2}sa, y
        + 2\sqrt{2}sb \rangle \}$; \\
        \lineno \> \> // construct four inequalities in the zero
        dimensional case\\
        \lineno \> \> if $Q=\{ \langle x_1, y_1 \rangle \}$ then
        return $\{ x \leq x_1, y \leq y_1, -x \leq -x_1, -y \leq
        -y_1\}$;\\
        \lineno \> \> // the centre of gravity $q_p$ is feasible
        but not a vertex (since $|Q|>1$)\\
        \lineno \> \> $q_p$ := $\langle \sum_{\langle x, y \rangle \in
          Q} x /|Q| , \sum_{\langle x, y \rangle \in Q} y /|Q|
        \rangle$;\\
        \lineno \> \> // $q_p$ is pivot point for sorting:
        $\forall i \in [0, n\!-\!2]$ . $\theta(\overline{q_p, q_i})
        \leq \theta(\overline{q_p, q_{i+1}})$\\
        \lineno \> \> $\seq{q_0,\ldots, q_{n-1}}$ :=
        $\mathrm{sort}(q_p, Q)$ \\
        \lineno \> \> // identify the $m$ vertices $q_{k_i}$ where 
        $0 \leq k_0 < \ldots < k_{m-1} < n$\\
        \lineno \> \> $\seq{ q_{k_0},\ldots, q_{k_{m-1}} }$ :=
        $\mathrm{scan}(\seq{q_0,\ldots, q_{n-1}})$\\
        \lineno \> \> $\res$ := $\emptyset$;\\
        \lineno \> \> for $i \in [0, m-1]$ do begin\\
        \lineno \> \> \> let $\langle x_1,
        y_1 \rangle = q_{k_i}$, $\langle x_2, y_2 \rangle =
        q_{k_{(i+1)\; mod \;m}}$\\
        \lineno \> \> \> let $e = \con(\langle x_1,
        y_1 \rangle, \langle x_2, y_2 \rangle)$\\
        \lineno \> \> \> // add $e$ to $\res$ if $q_{k_i}$ or
        $q_{k_{(i+1)\; mod \;m}}$ is in the box...\\
        \lineno \> \> \> $\add := \inBox(s,\langle x_1, y_1 \rangle)
        \vee \inBox(s,\langle x_2, y_2 \rangle) \vee m=2$;\\
        \lineno \> \> \> $j:=(k_i+1) \; \mod \; n$;\\
        \lineno \> \> \> while $\neg \add \wedge j \neq k_{i+1}$ do
        begin\\
        \lineno \> \> \> \> // ...or any boundary point is in the box\\
        \lineno \> \> \> \> $\add := \saturates(q_j, e) \wedge
        \inBox(s,q_j)$;\\
        \lineno \> \> \> \> $j:=(j+1) \; \mod \; n$;\\
        \lineno \> \> \> end;\\
        \lineno \> \> \> if $m=2 \wedge \inBox(s, \langle x_1,
        y_1 \rangle)$ then\\
        \lineno \> \> \> \> if $y_1=y_2$ \= then \= $\res\! :=
        \!\res \cup \{ \sgn(x_1-\!x_2) x \leq \sgn(x_1\!-\!x_2) x_1 \}$\\
        \lineno \> \> \> \> \> else \> $\res\! := \!\res \cup \{
        \sgn(y_1-y_2) y \leq \sgn(y_1-y_2) y_1 \}$\\
        \lineno \> \> \> if $\add$ then $\res$ := $\res \cup
        \{e\}$;\\
        \lineno \> \> \> end\\
        \lineno \> \> end\\
        \lineno \> \> return $\res$ \\
        \lineno \> end
      \end{tabbing}
    \end{minipage}}
    \caption{Convex hull algorithm for planar
      polyhedra}\label{fig-hull}
\end{figure}

The algorithm divides into a decomposition and a reconstruction phase.
The $\hull$ function decomposes the input polyhedra into their
corresponding ray and vertex representations by calling the function
$\extreme$ in lines 3 and 4. The remainder of the $\hull$ function
reconstructs a set of inequalities whose half\-spaces enclose both
sets of rays and points.  The functions $\extreme$ and $\hull$ are
presented in Figures \ref{fig-aux} and \ref{fig-hull}, respectively.
The algorithm requires the input polyhedra to be non-redundant. This
means that no proper subset of the inequalities induces the same space
as the original set of inequalities. The algorithm itself produces a
non-redundant system.

\newsavebox{\cross}

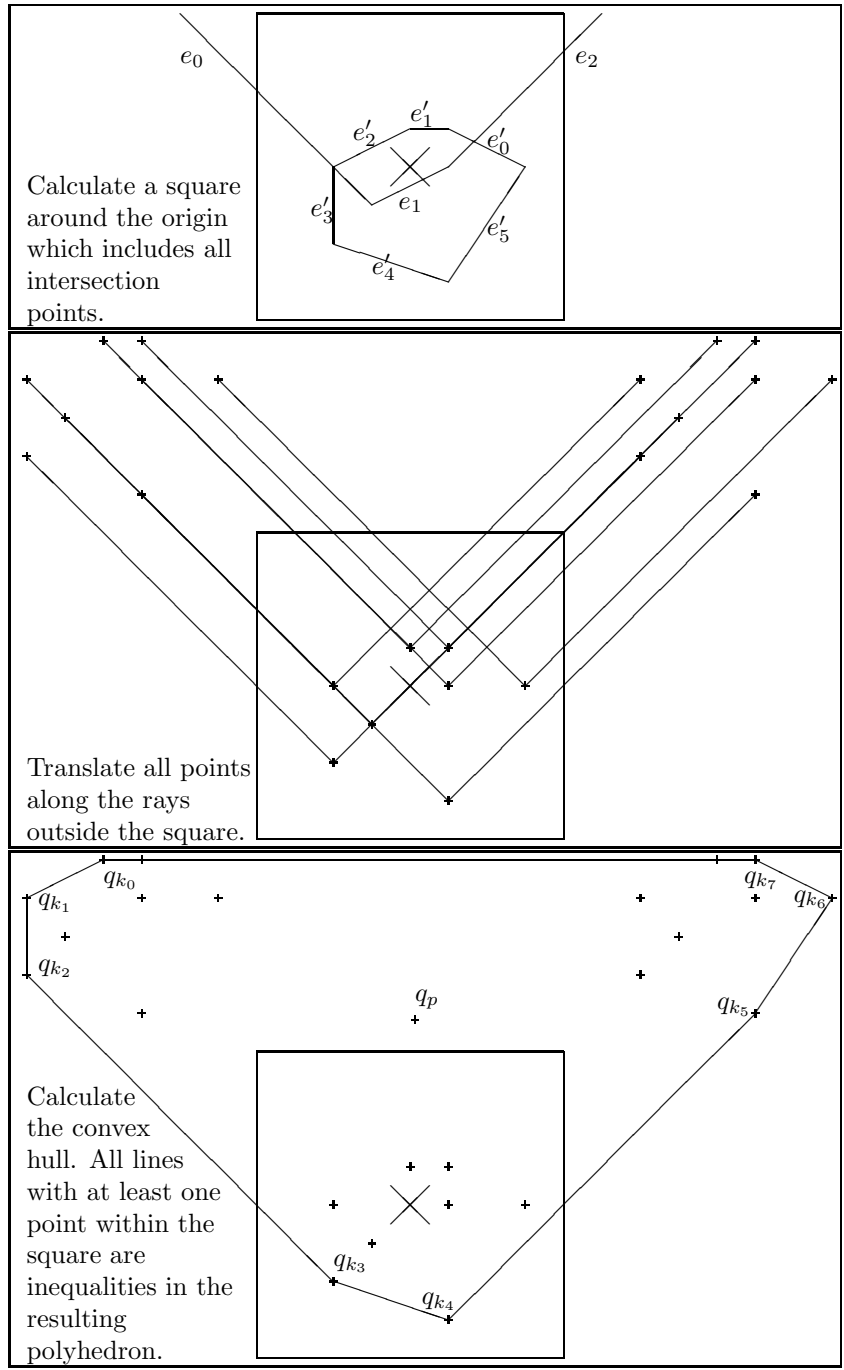
\begin{figure}
  \begin{center}
    \setlength{\unitlength}{0.51cm}
    \savebox{\cross}(0,0){
      \put(0.1,0){\line(-1,0){0.2}}
      \put(0,0.1){\line(0,-1){0.2}}
    }
    \fbox{
      \begin{picture}(21,8)
        \put(9.5,3.5){\line(1,1){1}}
        \put(9.5,4.5){\line(1,-1){1}}
        \put(6,0){\line(1,0){8}}
        \put(6,8){\line(1,0){8}}
        \put(6,0){\line(0,1){8}}
        \put(14,0){\line(0,1){8}}
        \put(11,1){\line(2,3){2}}
        \put(12,3){\parbox[t][0.4cm][b]{0cm}{$e'_5$}}
        \put(13,4){\line(-2,1){2}}
        \put(12,5.3){\parbox[t][0.4cm][b]{0cm}{$e'_0$}}
        \put(11,5){\line(-1,0){1}}
        \put(10,6){\parbox[t][0.4cm][b]{0cm}{$e'_1$}}
        \put(10,5){\line(-2,-1){2}}
        \put(8.5,5.5){\parbox[t][0.4cm][b]{0cm}{$e'_2$}}
        \put(8,4){\line(0,-1){2}}
        \put(7.4,3.5){\parbox[t][0.4cm][b]{0cm}{$e'_3$}}
        \put(8,2){\line(3,-1){3}}
        \put(9,2){\parbox[t][0.4cm][b]{0cm}{$e'_4$}}
        \put(4,8){\line(1,-1){5}}
        \put(4,7.5){\parbox[t][0.4cm][b]{0cm}{$e_0$}}
        \put(9,3){\line(2,1){2}}
        \put(9.7,3.7){\parbox[t][0.4cm][b]{0cm}{$e_1$}}
        \put(11,4){\line(1,1){4}}
        \put(14.3,7.5){\parbox[t][0.4cm][b]{0cm}{$e_2$}}
        \put(0,0){\parbox[b]{2.9cm}{
            \begin{flushleft}
              Calculate a square around the origin which includes all
              intersection points.
            \end{flushleft}\vspace{-\baselineskip}}}

      \end{picture}}
    \vfill
    \fbox{
      \begin{picture}(21,13)
        \put(9.5,3.5){\line(1,1){1}}
        \put(9.5,4.5){\line(1,-1){1}}
        \put(6,0){\line(1,0){8}}
        \put(6,8){\line(1,0){8}}
        \put(6,0){\line(0,1){8}}
        \put(14,0){\line(0,1){8}}
        \put(11,1){\usebox{\cross}}
        \put(13,4){\usebox{\cross}}
        \put(11,5){\usebox{\cross}}
        \put(10,5){\usebox{\cross}}
        \put(8,4){\usebox{\cross}}
        \put(8,2){\usebox{\cross}}
        \put(9,3){\usebox{\cross}}
        \put(11,4){\usebox{\cross}}
        \put(3,9){\usebox{\cross}}
        \put(5,12){\usebox{\cross}}
        \put(3,13){\usebox{\cross}}
        \put(2,13){\usebox{\cross}}
        \put(0,12){\usebox{\cross}}
        \put(0,10){\usebox{\cross}}
        \put(1,11){\usebox{\cross}}
        \put(3,12){\usebox{\cross}}
        \put(19,9){\usebox{\cross}}
        \put(21,12){\usebox{\cross}}
        \put(19,13){\usebox{\cross}}
        \put(18,13){\usebox{\cross}}
        \put(16,12){\usebox{\cross}}
        \put(16,10){\usebox{\cross}}
        \put(17,11){\usebox{\cross}}
        \put(19,12){\usebox{\cross}}
        \put(11,1){\line(-1,1){8}}
        \put(13,4){\line(-1,1){8}}
        \put(11,5){\line(-1,1){8}}
        \put(10,5){\line(-1,1){8}}
        \put(8,4){\line(-1,1){8}}
        \put(8,2){\line(-1,1){8}}
        \put(9,3){\line(-1,1){8}}
        \put(11,4){\line(-1,1){8}}
        \put(11,1){\line(1,1){8}}
        \put(13,4){\line(1,1){8}}
        \put(11,5){\line(1,1){8}}
        \put(10,5){\line(1,1){8}}
        \put(8,4){\line(1,1){8}}
        \put(8,2){\line(1,1){8}}
        \put(9,3){\line(1,1){8}}
        \put(11,4){\line(1,1){8}}
        \put(0,0){\parbox[b]{3cm}{
            \begin{flushleft}
              Translate all points along the rays outside the square.
            \end{flushleft}\vspace{-\baselineskip}}}
      \end{picture}}
    \vfill
    \fbox{
      \begin{picture}(21,13)
        \put(9.5,3.5){\line(1,1){1}}
        \put(9.5,4.5){\line(1,-1){1}}
        \put(6,0){\line(1,0){8}}
        \put(6,8){\line(1,0){8}}
        \put(6,0){\line(0,1){8}}
        \put(14,0){\line(0,1){8}}
        \put(11,1){\usebox{\cross}}
        \put(13,4){\usebox{\cross}}
        \put(11,5){\usebox{\cross}}
        \put(10,5){\usebox{\cross}}
        \put(8,4){\usebox{\cross}}
        \put(8,2){\usebox{\cross}}
        \put(9,3){\usebox{\cross}}
        \put(11,4){\usebox{\cross}}
        \put(3,9){\usebox{\cross}}
        \put(5,12){\usebox{\cross}}
        \put(3,13){\usebox{\cross}}
        \put(2,13){\usebox{\cross}}
        \put(0,12){\usebox{\cross}}
        \put(0,10){\usebox{\cross}}
        \put(1,11){\usebox{\cross}}
        \put(3,12){\usebox{\cross}}
        \put(19,9){\usebox{\cross}}
        \put(21,12){\usebox{\cross}}
        \put(19,13){\usebox{\cross}}
        \put(18,13){\usebox{\cross}}
        \put(16,12){\usebox{\cross}}
        \put(16,10){\usebox{\cross}}
        \put(17,11){\usebox{\cross}}
        \put(19,12){\usebox{\cross}}
        \put(10.125,8.833){\usebox{\cross}}
        \put(10.125,8.833){\parbox[b][0.4cm][l]{0.0cm}{$q_p$}}
        \put(11,1){\line(1,1){8}}
        \put(10.3,1.3){\parbox{0cm}{$q_{k_4}$}}
        \put(19,9){\line(2,3){2}}
        \put(18,9){\parbox{0.0cm}{$q_{k_5}$}}
        \put(21,12){\line(-2,1){2}}
        \put(20,11.8){\parbox{0.0cm}{$q_{k_6}$}}
        \put(19,13){\line(-1,0){17}}
        \put(18.7,12.3){\parbox{0.0cm}{$q_{k_7}$}}
        \put(2,13){\line(-2,-1){2}}
        \put(2,12.3){\parbox{0.0cm}{$q_{k_0}$}}
        \put(0,12){\line(0,-1){2}}
        \put(0.3,11.7){\parbox{0.0cm}{$q_{k_1}$}}
        \put(0,10){\line(1,-1){8}}
        \put(0.3,10){\parbox{0.0cm}{$q_{k_2}$}}
        \put(8,2){\line(3,-1){3}}
        \put(8,2.3){\parbox{0.0cm}{$q_{k_3}$}}
        \put(0,0){\parbox[b]{3.0cm}{
            \begin{flushleft}
              Calculate \linebreak the convex \linebreak hull. All
              lines \linebreak with at least one point within the
              square are inequalities in the resulting polyhedron.
            \end{flushleft}\vspace{-\baselineskip}}}
      \end{picture}}
    \caption{The different stages of the polyhedra convex hull
      algorithm.}
    \label{fig-hull-example1}
  \end{center}
\end{figure}

To illustrate the algorithm consider Figure \ref{fig-hull-example1}.
The polyhedron $E = \{ e_0, e_1, e_2 \}$ and the polytope $E' = \{
e'_0, \ldots, e'_5 \}$ constitute the input to the $\hull$ function.
They are passed to the function $\extreme$ at line 28 and 29.  Within
$\extreme$ the inequalities of each polyhedron are sorted at line 2.
Note that for ease of presentation the indices coincide with the
angular ordering.  The loop at lines 5--13 examines the relationship
of each inequality with its two angular neighbours.  If $\postRay$ is
false, the intersection point $\intersect(e_i, \eplus)$ is a vertex
which is added at line 11.  Conversely, if $\postRay$ is true, the
intersection point is degenerate, that is, either $E$ contains a
single inequality or the angular difference between the current
inequality and its successor is greater or equal to $\pi$.  In the
example two vertices are created for $E$, namely $v_1$ and $v_2$ where
$\{ v_1 \} = \intersect(e_0,e_1)$ and $\{ v_2 \} =
\intersect(e_1,e_2)$.  The intersection point $\intersect(e_2,e_0)$ is
degenerate, thus it is not added to $V$; in fact the point lies
outside the feasible space.  Six vertices are created for $E'$.  Rays
are created at line 9 and 10 if the intersection point is degenerate.
The two rays along the boundaries of $e_i$ and $\eplus$ are generated
in loop iteration $i$ when $\postRay$ is true and iteration $(i+1) \;
\mod \; n$ when $\preRay$ is true.  In our example $\postRay$ is true
for $e_2$, generating a ray along the boundary of $e_2$ which recedes
in the direction of the first quadrant, whereas $\preRay$ is only true
for $e_0$ yielding a ray along $e_0$ which recedes towards the second
quadrant.  No rays are created for the polytope $E'$.

In
general both flags might be true. In this circumstance the current
inequality $e_i$ cannot define a vertex.  In this case an arbitrary
point on the boundary of the halfspace of $e_i$ is created at line 12
to fix its representing rays in space.  Another case not encountered
in this example arises when the polyhedron consists of a single
halfspace ($|E|=1$). In this case a third ray is created (line 4) to
indicate on which side the feasible space lies. Note that the maximum
number of elements in $R$ never exceeds four, which occurs when the
input defines two facing halfspaces.

The remainder of the $\hull$ function is dedicated to the
reconstruction phase.  The point and ray sets, returned by $\extreme$,
are merged at line 30 and 31.  At line 32 the size of a square is
calculated which includes all points in $P$.  The square has $\langle
s, s \rangle$, $\langle -s, s \rangle$, $\langle s, -s \rangle$,
$\langle -s, -s \rangle$ as its corners.  The square in the running
example is depicted in all three frames of Figure
\ref{fig-hull-example1} and the origin is marked with a cross.  Each
point $p \in P$ is then translated by each ray $r \in R$ yielding the
point set $Q$.  Translated points appear outside the square since all
normalised rays are translated by the length of the diagonal
$2\sqrt{2}s$ of the square.  The translation process for the worked
example is depicted in the second frame.  Line 38 is not relevant to
this example as it traps the case when the output polyhedron consists
of a single point.  Line 40 calculates a feasible point $q_p$ of the
convex hull of $Q$ which is not a vertex.  This point serves as the
pivot point in the classic Graham scan.  First, the point set $Q$ is
sorted counter-clockwise with respect to $q_p$.  Second, interior
points are removed, yielding the indices of all vertices, in the case
of the example $k_0, \ldots, k_7$.  What follows is a round-trip
around the hull which translates pairs of adjacent vertices into
inequalities by calling $\con$ at line 48.  Whether this inequality
actually appears in the result depends on the state of the $\add$
flag.  In our particular example the $\add$ flag is only set at line
50.  Whenever it is set, it is because one of the two vertices lies
within the square.  The resulting polyhedron consists of the
inequalities $\overline{q_{k_2}, q_{k_3}}$, $\overline{q_{k_3},
  q_{k_4}}$ and $\overline{q_{k_4}, q_{k_5}}$ which is a correct
solution for this example.
 
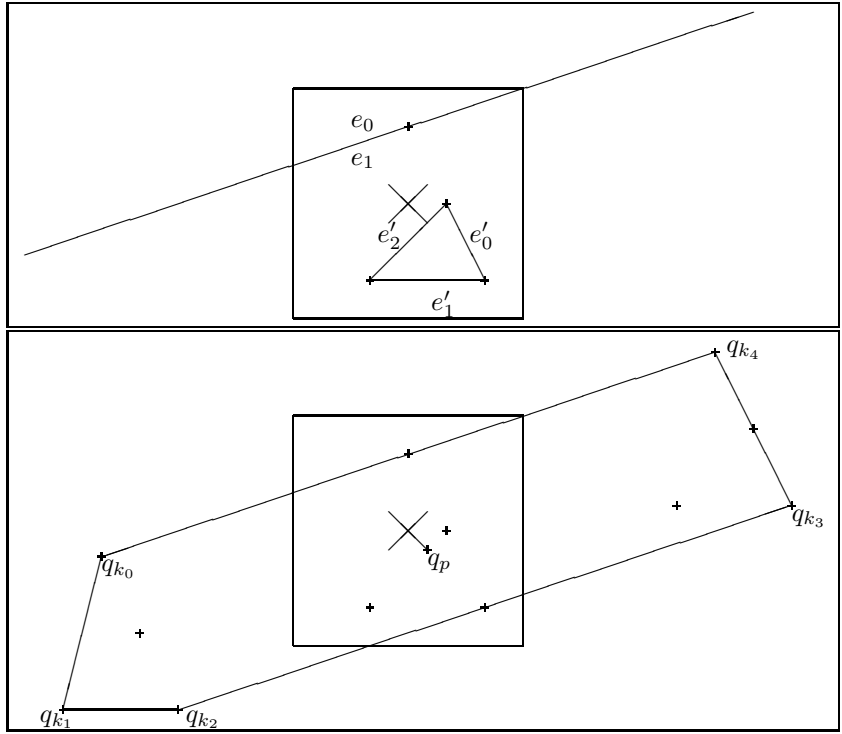
\begin{figure}[t]
  \begin{center}
    \setlength{\unitlength}{0.51cm}
    \savebox{\cross}(0,0){
      \put(0.1,0){\line(-1,0){0.2}}
      \put(0,0.1){\line(0,-1){0.2}}
    }
    \fbox{
      \begin{picture}(21,8)
        \put(9.5,2.5){\line(1,1){1}}
        \put(9.5,3.5){\line(1,-1){1}}
        \put(7,0){\line(1,0){6}}
        \put(7,6){\line(1,0){6}}
        \put(7,0){\line(0,1){6}}
        \put(13,0){\line(0,1){6}}
        \put(9,1){\usebox{\cross}}
        \put(9,1){\line(1,1){2}}
        \put(11.6,2){\parbox[b][0.4cm][b]{0cm}{$e'_0$}}
        \put(12,1){\usebox{\cross}}
        \put(12,1){\line(-1,0){3}}
        \put(10.6,1){\parbox[t][0.4cm][b]{0cm}{$e'_1$}}
        \put(11,3){\usebox{\cross}}
        \put(11,3){\line(1,-2){1}}
        \put(9.2,2){\parbox[b][0.4cm][b]{0cm}{$e'_2$}}
        \put(10,5){\usebox{\cross}}
        \put(0,1.666){\line(3,1){19}}
        \put(8.5,5){\parbox[b][0.4cm][b]{0cm}{$e_0$}}
        \put(8.5,4.8){\parbox[t][0.4cm][b]{0cm}{$e_1$}}
      \end{picture}}
    \vfill
    \fbox{
      \begin{picture}(21,10)
        \put(9.5,4.5){\line(1,1){1}}
        \put(9.5,5.5){\line(1,-1){1}}
        \put(7,2){\line(1,0){6}}
        \put(7,8){\line(1,0){6}}
        \put(7,2){\line(0,1){6}}
        \put(13,2){\line(0,1){6}}
        \put(9,3){\usebox{\cross}}
        \put(12,3){\usebox{\cross}}
        \put(11,5){\usebox{\cross}}
        \put(1,0.33){\usebox{\cross}}
        \put(4,0.33){\usebox{\cross}}
        \put(3,2.33){\usebox{\cross}}
        \put(17,5.66){\usebox{\cross}}
        \put(20,5.66){\usebox{\cross}}
        \put(19,7.66){\usebox{\cross}}
        \put(10,7){\usebox{\cross}}
        \put(2,4.33){\usebox{\cross}}
        \put(18,9.66){\usebox{\cross}}
        \put(20,5.66){\line(-1,2){2}}
        \put(18,9.66){\line(-3,-1){16}}
        \put(2,4.33){\line(-1,-4){1}}
        \put(1,0.33){\line(1,0){3}}
        \put(4,0.33){\line(3,1){16}}
        \put(10.5,4.5){\usebox{\cross}}
        \put(10.5,4.5){\parbox[t][0.2cm][b]{0cm}{$q_p$}}
        \put(2,4.33){\parbox[t][0.2cm][b]{0cm}{$q_{k_0}$}}
        \put(0.4,0.4){\parbox[t][0.2cm][b]{0cm}{$q_{k_1}$}}
        \put(4.2,0.4){\parbox[t][0.2cm][b]{0cm}{$q_{k_2}$}}
        \put(20,5.66){\parbox[t][0.2cm][b]{0cm}{$q_{k_3}$}}
        \put(18.3,9.66){\parbox[b][0.2cm][b]{0cm}{$q_{k_4}$}}
      \end{picture}}
    \caption{The resulting convex hull contains a line.}
    \label{fig-hull-example3}
  \end{center}
\end{figure}

In general, the reconstruction phase has to consider certain anomalies
that mainly arise in outputs of lower dimensionality.  One subtlety in
the two dimensional case is the handling of polyhedra which contain
lines.  This is illustrated in Figure \ref{fig-hull-example3} where
the two inequalities $e_0, e_1$ are equivalent to one equation which
defines a space that is a line.  Observe from the second frame that no
point in the square is a vertex in the hull of $Q$.  Therefore the
predicate $\inBox$ does not hold for the two vertices $q_{k_2}$ and
$q_{k_3}$ and the desired inequality $\overline{q_{k_2}, q_{k_3}}$ is
not emitted.  Similarly for the vertices $q_{k_4}$ and $q_{k_0}$.
However, in such cases there always exists a point in $p \in Q$ with
$\ang{\overline{q_p, q_{k_i}}}{\overline{q_p, p}} <
\ang{\overline{q_p, q_{k_i}}}{\overline{q_p, q_{k_{(i+1) \; \mod \;
        m}}}}$ which lies in the square.  Thus it is sufficient to
search for an index $j \in [k_{i}+1, k_{(i+1) \; \mod \; m}-1]$ such
that $q_j$ is both in the square and on the line connecting the
vertices $q_{k_i}$ and $q_{k_{(i+1) \; \mod \; m}}$.  The inner loop
at lines 52--56 tests if $Q$ contains such a point and sets $\add$
appropriately.

\begin{figure}
  \begin{center}
    \setlength{\unitlength}{0.6cm}
    \savebox{\cross}(0,0){
      \put(0.1,0){\line(-1,0){0.2}}
      \put(0,0.1){\line(0,-1){0.2}}
    }
    \fbox{
      \begin{picture}(14,7)
        \put(6.5,3.0){\line(1,1){1}}
        \put(6.5,4.0){\line(1,-1){1}}
        \put(5,1.5){\line(1,0){4}}
        \put(5,5.5){\line(1,0){4}}
        \put(5,1.5){\line(0,1){4}}
        \put(9,1.5){\line(0,1){4}}
        \put(2,1.5){\line(2,1){10}}
        \put(12,6.5){\usebox{\cross}}
        \put(11,6){\usebox{\cross}}
        \put(10,5.5){\usebox{\cross}}
        \put(8,4.5){\usebox{\cross}}
        \put(7,4){\usebox{\cross}}
        \put(6,3.5){\usebox{\cross}}
        \put(4,2.5){\usebox{\cross}}
        \put(3,2){\usebox{\cross}}
        \put(2,1.5){\usebox{\cross}}
        \put(12,6.5){\parbox[t][0.4cm][b]{0cm}{$q_8$}}
        \put(11,6){\parbox[b][0.4cm][t]{0cm}{$q_7$}}
        \put(10,5.5){\parbox[b][0.4cm][t]{0cm}{$q_6$}}
        \put(8,4.5){\parbox[b][0.4cm][t]{0cm}{$q_5$}}
        \put(7,4){\parbox[b][0.4cm][t]{0cm}{$q_4$}}
        \put(6,3.5){\parbox[b][0.4cm][t]{0cm}{$q_3$}}
        \put(4,2.5){\parbox[b][0.4cm][t]{0cm}{$q_2$}}
        \put(3,2){\parbox[b][0.4cm][t]{0cm}{$q_1$}}
        \put(2,1.5){\parbox[b][0.4cm][t]{0cm}{$q_0$}}
        \put(7,0.8){\line(0,1){0.4}}
        \put(7,1){\line(1,0){2}}
        \put(7.8,1){\parbox[t][0.4cm][b]{0cm}{$s$}}
      \end{picture}}
    \caption{Handling the one-dimensional case.}
    \label{fig-hull-example2}
  \end{center}
\end{figure}
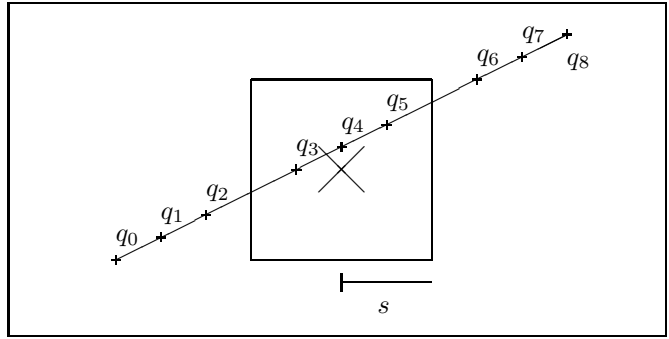

The one dimensional case is handled by the $m=2$ tests at line 50 and
57.  Figure \ref{fig-hull-example2} illustrates why the test in line
50 is necessary.  Suppose $E_1$ and $E_2$ are given such that
$\extreme(E_1) = \langle \{ q_4, q_5 \}, \emptyset \rangle$ and
$\extreme(E_2) = \langle \{ q_3 \}, \{ r, -r \} \rangle$ where $r$ is
any ray parallel to the line.  Observe that all points are collinear,
thus the pivot point is on the line and a stable sort could return the
ordering depicted in the figure.  The correct inequalities for this
example are $\res = \{ \overline{q_0, q_8}, \overline{q_8, q_0} \}$.
The Graham scan will identify $q_{k_0}=q_0$ and $q_{k_1}=q_8$ as
vertices. Since there exists $j \in [k_0+1, k_1-1]$ such that
$\inBox(s,q_j)$ holds, $\overline{q_0, q_8} \in \res$. In contrast,
although there are boundary points between $q_8$ and $q_0$ the loop
cannot locate them due to the ordering.  In this case the $m=2$ test
ensures that $\add$ is set, guaranteeing that $\overline{q_8, q_0} \in
\res$.

The output polyhedron must include $q_{k_i}$ as a vertex whenever
$\inBox(s, q_{k_i})$ holds.  If $\inBox(s, q_{k_i})$ holds, the
algorithm generates $e_{i-1} = \overline{q_{k_{(i-1) \; \mod \; m}},
  q_{k_i}}$ and $e_i = \overline{q_{k_i}, q_{k_{(i+1) \; \mod \;
      m}}}$. If $\ang{e_{i-1}}{e_i} < \pi$, then $\{ q_{k_i} \} =
\intersect(e_{i-1}, e_i)$ and the vertex $q_{k_i}$ is realized.
Observe that if $m=2$, $\ang{e_{i-1}}{e_i} = \pi$ which necessitates
an additional inequality to define the vertex $q_{k_i}$.  This is the
r\^ole of the inequality generated on lines 58 or 59.  Observe that
this inequality $e$ guarantees $\ang{e_{i-1}}{e} < \pi$ and
$\ang{e}{e_i} < \pi$ which is sufficient to define $q_{k_i}$.

The zero dimensional case corresponds to the case of when both input
polyhedra consist of the single point $v$.  Line 38 traps this case
and returns a set of inequalities describing $\{ v \}$. Observe that
the zero and one dimensional case only require minute changes to the
general two dimensional case.

As a note on implementation, observe that the search for a pivot point
at line 40 can be refined.  One method for finding a definite vertex
is to search for a point with extremal coordinates
\cite{anderson87reevaluation}. However, this process requires all
points to be examined. The presented algorithm follows Graham
\cite{graham72efficient} in creating an interior point as the pivot
point. This does not necessarily require the whole point set to be
examined.  By choosing two arbitrary points $q_1, q_2$, it is
sufficient to search the point set for a point $q_i$ which does not
saturate the line $\overline{q_1, q_2}$.  The centre of the triangle
$q_1, q_2, q_i$ is guaranteed to be an interior point of $Q$.

Note also that the sorting in $\extreme$ is unnecessary if the input
inequalities are consecutive in terms of angle. In particular, the
output of one run of the algorithm can serve as an input to another
without applying the sort at line 2.

Finally observe that the inner loop at lines 52--56 can often be
skipped: if the line between $q_{k_i}$ and $q_{k_{(i+1) \; \mod \;
    m}}$ does not intersect with the square, $\inBox(s,q)$ cannot hold
for any $q \in Q$. Hence $\add$ will not be set at line 54 and the
inner loop has no effect. 

\section{\label{sec-correct}Proof of Correctness}

Section \ref{sec-prelim} introduces the mathematical language
necessary for expressing the two parts of the proof.  The proof itself
reflects the structure of the algorithm: Section \ref{sec-extreme}
concerns the conversion of planar $H$-polyhedra into their ray and
point representations; Section \ref{sec-reconstruct} argues that the
reconstructed polyhedron encloses the points and rays generated from
the two input polyhedra, and yet is also minimal.

\subsection{\label{sec-prelim}Preliminaries}

A polyhedron is a set $P \subseteq \real^d$ such that $P = \{ \vec{x}
\in \real^d \mid A\vec{x} \leq \vec{c} \}$ for some matrix $A \in
\real^{n \times d}$ and vector $\vec{c} \in \real^n$.  An
\mbox{H-polyhedron} is a pair $\langle A, \vec{c} \rangle$ where $A
\in \real^{n \times d}$ and $\vec{c} \in \real^n$ that is interpreted
by $\lbag . \rbag$ as the polyhedron $\lbag \langle A, \vec{c} \rangle
\rbag = \{ \vec{x} \in \real^d \mid A\vec{x} \leq \vec{c} \}$.  For
brevity we manipulate $H$-polyhedra as a finite set of inequalities $E
= \{ \vec{a}_1 \cdot \vec{x} \leq c_1, \ldots, \vec{a}_n \cdot \vec{x}
\leq c_n \}$ which is equivalent to the matrix representation $\langle
A, \vec{c} \rangle$ with $A = \langle \vec{a}_1, \ldots, \vec{a}_n
\rangle^T$ and $\vec{c}= \langle c_1, \ldots, c_n \rangle^T$.  $E$ is
said to be satisfiable if $\lbag E \rbag \neq \emptyset$ and
non-redundant if $\forall e \in E\; .\; \lbag E \setminus \{e\} \rbag
\neq \lbag E \rbag$. Two inequalities $e, e'$ coincide, written
$\coincide(e, e')$, iff there exists $\vec{a} \in \real^d, c \in
\real$ with $\lbag \{ e, e' \} \rbag = \{ \vec{x} \in \real^d \mid
\vec{a} \cdot\vec{x} = c \}$.

The convex hull of a finite set of points $P = \{ \vec{p}_1, \ldots,
\vec{p}_n \} \subseteq \real^d$ is defined as $\conv(P) = \{
\sum_{i=1}^{n} \lambda_i \vec{p}_i \mid 0 \leq \lambda_i \wedge
\sum_{i=1}^{n} \lambda_i = 1 \}$.  Moreover, the cone of a finite set
of vectors $R = \{ \vec{r}_1, \ldots, \vec{r}_m \} \subseteq \real^d$
is defined as $\cone(R) = \{ \sum_{i=1}^{m} \lambda_i \vec{r}_i \mid 0
\leq \lambda_i \}$.  The Minkowski sum of two sets $X, Y \subseteq
\real^d$ is defined as $X+Y = \{ \vec{x} + \vec{y} \mid \vec{x} \in X
\wedge \vec{y} \in Y \}$.  Let $\ext(S) = \{ p \in S \mid p \notin
\conv(S \setminus \{ p \})\}$, $\ray(S) = \{ r \in \real^d \setminus
\{ \vec{0} \} \mid S + \cone(\{ r \}) \subseteq S \}$ and $\dray(S) =
\{ r \in \ray(S) \mid -r \in \ray(S) \}$ denote the vertices, rays and
lines of a convex set $S$.  The following result, accredited to
Motzkin \cite{ziegler}, relates the two classic representation of
polyhedra.

\begin{theorem} \label{theo-equiv} \rm
  The following statements are equivalent for any $S \subseteq
  \real^d$:
  \begin{enumerate}
  \item\label{case-a} $S = \conv(P) + \cone(R)$ for some finite $P, R
    \subseteq \real^d$;
  \item\label{case-b} $S = \{ \vec{x} \in \real^d \mid A\vec{x} \leq
    \vec{c} \}$ for some matrix $A \in \real^{n \times d}$ and vector
    $\vec{c} \in \real^n$.
  \end{enumerate}
\end{theorem}

Our algorithm converts the two (planar) input $H$-polyhedra $E_i$ into
their rays $R_i$ and points $P_i$ and calculates an $H$-polyhedron $E$
with the smallest $\lbag E \rbag$ such that $\lbag E \rbag \supseteq
\lbag E_1 \rbag \cup \lbag E_2 \rbag$. In fact $\lbag E \rbag =
\conv(P_1 \cup P_2) + \cone(R_1 \cup R_2)$. 

\subsection{\label{sec-extreme}Decomposition}
The discussion of the function $\extreme$ is organised by the
dimension of the polyhedron. Reoccurring or self-contained arguments
are factored out in the following lemmata.  The first lemma states how
redundancy can follow from the angular relationship between three
inequalities.  Since $\extreme$ requires its input to be
non-redundant, useful angular properties of the input inequalities
flow from the lemma.

\begin{lemma}\label{lemAngle}
  Suppose $e_i = a_i x + b_i y \leq c_i$, $i = 1,2,3$. Let $c_1= c_3
  =0$, $c_2 \geq 0$. Then $\lbag \{ e_1, e_3 \} \rbag \subseteq \lbag
  \{ e_2 \} \rbag$ if $0 = \theta(e_1) < \theta(e_2) < \theta(e_3) <
  \pi$.
\end{lemma}

\begin{proof}
  To show $\lbag \{ e_1, e_3 \} \rbag \subseteq \lbag \{ e_2 \} \rbag$
  it is sufficient to show $\lbag \{ e_1, e_3 \} \rbag \subseteq \lbag
  \{ a_2 x + b_2 y \leq 0 \} \rbag$, thus let $c_2 = 0$.  Furthermore,
  w.l.o.g.~let $e_1 \equiv x \leq 0$ (since $\theta(e_1) = 0$), $e_2
  \equiv a_2 x + y \leq 0$ (since $\theta(e_2) < \pi$) and $e_3 \equiv
  a_3 x + y \leq 0$ (since $\theta(e_3) < \pi$). Note that $a_2 =
  \lambda_1 a_1 + \lambda_3 a_3$ and $b_2 = \lambda_1 b_1 + \lambda_3
  b_3$ has the solution $\lambda_1 = a_2 - a_3$ and $\lambda_3=1$.
  Due to $b_2=1$, $a_2 = \cot^{-1}(\theta(e_2))$ and similarly $a_3 =
  \cot^{-1}(\theta(e_3))$.  Because $\theta(e_2) < \theta(e_3)$ and
  $\cot$ is an anti-monotone on $(0,\pi)$, it follows that $a_2 >
  a_3$, hence $\lambda_1 >0$.  Let $\langle x, y \rangle$ satisfy $e_1
  \equiv x \leq 0$ and $e_3 \equiv a_3 x + y \leq 0$. Due to
  $\lambda_1 x \leq 0$, $\lambda_1(x) + \lambda_3(a_3 x + y) =
  (\lambda_1 + a_3)x + y \leq 0 \equiv e_2$, thus $e_2$ holds, hence
  $\lbag \{ e_1, e_3 \} \rbag \subseteq \lbag \{ e_2 \} \rbag$ as
  required.
\end{proof}

\noindent The following lemma states that there is an injection
between vertices and inequalities. This result is used to show that
the loop in $\extreme$ does indeed generate all vertices of the input
polyhedron.

\begin{lemma} \label{lemInjection}
  Let $E = \{e_0, \ldots, e_{n-1}\}$ be non-redundant and ordered by
  $\theta$ and $v \in \ext(\lbag E \rbag)$. Then there exists $e_i \in
  E$ such that $\{v\} = \intersect(e_i, e_{i'})$ and
  $\ang{e_i}{e_{i'}} < \pi$ where $i' = (i+1) \; \mod \; n$.
\end{lemma}

\begin{proof} Let $E' \subseteq E$ contain those inequalities that $v$
  saturates. Note that $|E'| \geq 2$, in particular there exist $e,e'
  \in E'$ with $\ang{e}{e'} \notin \{ 0, \pi \}$, otherwise $\lbag E'
  \rbag \setminus \{v\}$ is not convex.  Choose $e_i, e_k \in E'$ such
  that $\ang{e_i}{e_k} < \pi$.  Suppose for the sake of a
  contradiction that $k \neq (i+1) \; \mod \; n$.  Then $e_j \in E
  \setminus E'$ exists with $\ang{e_i}{e_j} < \ang{e_i}{e_k}$.
  W.l.o.g.~assume that $\theta(e_i)=0$ and $v = \langle 0, 0 \rangle$.
  Then $\ang{e_i}{e_j} < \ang{e_i}{e_k}$ reduces to $0 = \theta(e_i) <
  \theta(e_j) < \theta(e_k) < \pi$.  Lemma \ref{lemAngle} implies
  $\lbag E \setminus \{ e_j \} \rbag \subseteq \lbag E \rbag$ which
  contradicts the assumption that $E$ has no redundant inequalities.
  Thus $k = (i+1) \; \mod \; n = i'$
\end{proof}

\noindent 
The following result also builds on Lemma \ref{lemAngle} and
complements the previous lemma in that it states when 
 adjacent inequalities have
feasible intersections points. The lemma is used to show that 
$\extreme$ only generates points in $\lbag E \rbag$.

\begin{lemma} \label{lemIntersect}
  Let $E = \{e_0, \ldots, e_{n-1}\}$ be satisfiable, non-redundant and
  ordered by $\theta$. For any $e_i \in E$ if $\ang{e_i}{\eplus}
  < \pi$ then $\intersect(e_i, \eplus) \subseteq \lbag E \rbag$.
\end{lemma}

\begin{proof}
  Let $e_i \in E$. Since $e_i$ is not redundant in $E$ the boundary of
  $e_i$ intersects with the non-empty convex body $\lbag E \setminus
  \{ e_i \} \rbag$.  Let $e_m \in E \setminus \{ e_i \}$ such that
  $\emptyset \neq S = \intersect(e_i, e_m) \subseteq \lbag E\setminus
  \{ e_i \} \rbag$.  Note that if $|S| > 1$ then $E = \{ e_i, e_m\}$
  with $\ang{e_i}{e_m} = \pi$, hence $\ang{e_i}{\eplus} < \pi$ never
  holds.  Let $\{ v \} = S$.  It remains to show that $e_i$ and $e_m$
  are adjacent.  Suppose that $\ang{e_i}{e_m} < \pi$ ($\ang{e_m}{e_i}
  < \pi$ is analogous).  Assume for the sake of a contradiction there
  exists $e_l \in E \setminus \{ e_i \}$ such that $\ang{e_i}{e_l} <
  \ang{e_i}{e_m}$. W.l.o.g.~$v = \langle 0, 0 \rangle$ and
  $\theta(e_i)=0$, hence $0 = \theta(e_i) < \theta(e_l) < \theta(e_m)
  < \pi$. Since $v \in \lbag E \setminus \{ e_i \}\rbag$, $v \in \lbag
  e_l \rbag$ and thus $c_l \geq 0$ where $e_l \equiv a_l x + b_l y
  \leq c_l$. By Lemma \ref{lemAngle} $e_l$ is redundant in $E$ which
  is a contradiction. It follows that $m = (i+1) \; \mod \; n$.
\end{proof}

\noindent While the previous lemmata concern points, the following lemma is
a statement about rays.
In particular, it states when inequalities give rise to rays.

\begin{lemma} \label{lemRays}
  Let $E = \{ e_0 \equiv a_0 x + b_0 y \leq c_0, \ldots, e_{n-1} \}$
  be satisfiable, non-redundant and ordered by $\theta$.  Let
  $i'=(i+1) \; \mod \; n$, $S_1 = \cone(\{ \langle -b_i, a_i \rangle,
\linebreak
  \langle b_{i'}, -a_{i'} \rangle \})$ and $S_2 = \ray(\lbag E
  \rbag)$. If $\ang{e_i}{e_{i'}} > \pi$ or $|E| \geq 3 \wedge
  \ang{e_i}{e_{i'}} = \pi$ then $S_1 = S_2$.
\end{lemma}

\begin{proof}
  To show $S_1 \subseteq S_2$. Choose $p \in \lbag E \rbag$ such that
  $p$ saturates $e_i$. For the sake of a contradiction assume $r_i =
  \langle -b_i, a_i \rangle \notin \ray ( \lbag E \rbag )$.  Thus
  there exists $\lambda > 0$ with $p + \lambda r_i \notin \lbag E
  \rbag$.  Set $\lambda_{\max} = \max \{ \lambda >0 \mid p + \lambda
  r_i \in \lbag E \rbag \}$. Note that $v= p + \lambda_{\max} r_i$ is
  a vertex and by Lemma \ref{lemInjection} there exists $e_k \in E$
  with $\{ v \} = \intersect(e_k, e_{(k+1) \; \mod \; n})$ and
  $\ang{e_k}{e_{(k+1) \; \mod \; n}} < \pi$. The latter implies that
  $k \neq i$, therefore $i=(k+1) \; \mod \; n$.  Hence $e_{(i-1) \;
    \mod \; n} = e_k$ does not contain the ray.  W.l.o.g. let
  $\theta(e_i)=0$ and $v=\langle 0, 0 \rangle$. Then $r_i = \langle 0,
  1 \rangle$ and $\pi < \theta(e_{(i-1) \; \mod \; n}) < 2\pi$, hence
  $e_{(i-1) \; \mod \; n} \equiv a' x + b' y \leq 0$ with $b' < 0$.
  Thus $a' x + b' (\lambda + y) \leq 0$, hence $p + \lambda r_i \in
  \lbag \{ e_{(i-1) \; \mod \; n} \} \rbag$ which is a contradiction.
  Analogously for $r_{i'} = \langle b_{i'}, -a_{i'} \rangle$. Now to
  show $S_2 \subseteq S_1$.  Assume there exists $r \in \ray(\lbag E
  \rbag) \setminus \cone(\{ \langle -b_i, a_i \rangle, \langle b_{i'},
  -a_{i'} \rangle \})$.  Consider $\ang{e_i}{e_{i'}} > \pi$.  Since
  $\theta(e_i) \neq \theta(e_{i'})$ there exists $\lambda_1, \lambda_2
  \in \real$ with $r = \lambda_1 \langle -b_i, a_i \rangle + \lambda_2
  \langle b_{i'}, -a_{i'} \rangle$. Assume $\lambda_2 < 0$.  W.l.o.g.
  let $\{p\} = \intersect(e_i, e_{i'}) = \{ \langle 0, 0 \rangle \}$
  and $\theta(e_i) = 0$ thus let $e_i \equiv x \leq 0$.  It follows
  that $\pi < \theta(e_{i'}) < 2\pi$, $b_{i'} \leq 0$ and we set
  $e_{i'} \equiv a_{i'}' x -y \leq 0$.  Thus $r = \lambda_1 \langle 0,
  1 \rangle + \lambda_2 \langle -1, -{a_{i'}'} \rangle =\langle
  -\lambda_2, \lambda_1 - \lambda_2 a_{i'}' \rangle$.  Since
  $\lambda_2 < 0$, $p + r \notin \lbag \{ e_i \} \rbag$ which
  contradicts $r \in \ray(\lbag E \rbag)$.  Analogously for $\lambda_1
  <0$ with $\theta(e_{i'})=0$.  Since $\lambda_1 \geq 0$ and
  $\lambda_2 \geq 0$, $r \in \cone(\{ \langle -b_i, a_i \rangle,
  \langle b_{i'}, -a_{i'} \rangle \})$ which is a contradiction.  Now
  suppose $\ang{e_i}{e_{i'}} = \pi$ and $|E| \geq 3$.  W.l.o.g. let
  $\theta(e_i)=0$, $e_i \equiv x \leq |u|$ and $e_{i'} \equiv -x \leq
  0$.  Thus $S_1 = \cone(\{ \langle 0, 1 \rangle \})$.  Let $\langle
  x_r, y_r \rangle = r$ and $p \in \lbag \{ e_i, e_{i'} \} \rbag$.
  Observe that $x_r = 0$ otherwise there exists $\lambda >0$ with $p +
  \lambda r \notin \lbag \{ e_i, e_{i'} \} \rbag$.  Now assume $y_r <
  0$.  Since $|E| \geq 3$, there exists $e_j \in E$ such that $\pi <
  \theta(e_j) < 2\pi$ and thus $e_j \equiv a_j x + b_j y \leq c_j$
  with $b_j <0$.  Let $p \in \lbag \{ e_j \} \rbag$, then there exists
  $\lambda>0$ such that $p + \lambda r \notin \lbag \{ e_j \} \rbag$
  which is a contradiction.
\end{proof}

The correctness of the first stage of our algorithm is summarised by
the following proposition.  Note that the Minkowski sum $\conv(V) +
\emptyset$ always defines the empty space rather than the polyhedron
$\conv(V)$.  Thus a null ray $\langle 0, 0 \rangle$ is required to
represent a bounded polyhedron.  The function $\extreme$ avoids adding
this null ray for the sake of improved efficiency. In $\hull$ the
translation of points at line 36 by the null ray is replaced by a
simple copying step at line 35.  However, the null ray still manifests
itself in the correctness results.

\begin{proposition}
  Let $E \subset \Lin$ be non-empty, finite, satisfiable and
  non-re\-dun\-dant.  Then $\extreme(E) = \langle P, R \rangle$ and $\lbag
  E \rbag = \conv(P) + \cone(R \cup \{ \langle 0,0 \rangle \})$.
\end{proposition}

Note that the following proof handles polyhedra that contain lines as
a special case. This distinction is not artificial, in fact Klee
\cite{klee59some} observed that a closed convex set that does not
contain lines is generated by its vertices and its extreme rays.
(Extreme rays are rays that cannot be expressed by a linear
combination of others.)  In order to describe polyhedra which contain
lines it is necessary to create points that are not vertices and rays
which are not extreme.

\begin{proof}
  Let $\{e_0, \ldots, e_{n-1}\} =E$ such that $\theta(e_0) <
  \theta(e_1) < \ldots < \theta(e_{n-1})$. Such an ordering exists
  since if $\theta(e_i)=\theta(e_j)$ for some $i \neq j$ then either
  $\lbag e_i \rbag \subseteq \lbag e_j \rbag$, thus $\lbag E \setminus
  \{e_j\} \rbag = \lbag E \rbag$ or vice-versa which contradicts that
  $E$ is non-redundant. Let $\langle P, R \rangle = \extreme(E)$.
  \begin{itemize}
  \item Suppose $\dray(\lbag E \rbag) \neq \emptyset$.  Note that if
    there exist $e_i, e_j \in E$ with $\ang{e_i}{e_j} < \pi$ then
    $\dray(\lbag E \rbag) = \emptyset$, hence $E = \{ e\}$ or $E = \{
    e_0, e_1 \} \wedge \ang{e_0}{e_1} = \pi$.
    \begin{itemize}
    \item Consider $E = \{ e\}$. The loop is executed only once with
      $\preRay = \postRay = \true$, thus the boundary point $p$ and
      two rays $r_1, r_2$ are added in line 12, 9 and 10,
      respectively. The boundary of $\lbag E \rbag$ is $\{ p +
      \lambda_1 r_1 + \lambda_2 r_2 \mid \lambda_i \geq 0 \}$. The ray
      $r_3$ added in line 4 points into the halfspace, thus $\lbag E
      \rbag = \{ p \} + \cone(\{ r_1, r_2, r_3 \})$.
    \item Consider $E = \{ e_0, e_1 \} \wedge \ang{e_0}{e_1} = \pi$.
      For each $e_i$, $\preRay = \postRay = \true$, hence for each
      $e_i$ the loop generates two rays $r_i, r'_i$ along the boundary
      of $e_i$ (lines 9, 10) and one boundary point $p_i$ (line 11).
      The set $\conv(\{p_1, p_2\})$ is included in $\lbag E \rbag$.
      Note that the rays generated in the second iteration are
      collinear to those in the first.  Thus $\lbag E \rbag =
      \conv(\{p_1, p_2\}) + \cone(\{r_1, r'_1\})$.
    \end{itemize}
  \item Suppose $E$ is zero dimensional, hence $\lbag E \rbag = \{v
    \}$.  Let $\alpha_i = \ang{e_i}{\eplus}$.  Since $\sum_{i=0}^{n-1}
    \alpha_i = 2\pi$, it follows that $n \geq 3$ since otherwise there
    exists $i$ such that $\alpha_i \geq \pi$ and by Lemma
    \ref{lemRays} it follows that $\ray(\lbag E \rbag) \neq
    \emptyset$.  By Lemma \ref{lemAngle}, it follows that $\alpha_i +
    \alpha_{(i+1) \; \mod \; n} \geq \pi$, which for all $e_i$ is
    $ \sum_{i=0}^{n-1}
    (\alpha_i + \alpha_{(i+1) \; \mod \; n}) \geq n\pi$.  However, $n
    \pi \leq \sum_{i=0}^{n-1} (\alpha_i + \alpha_{(i+1) \; \mod \; n})
    = 2 \sum_{i=0}^{n-1} \alpha_i = 4\pi$, hence $n \leq 4$.
    \begin{itemize}
    \item Consider $E = \{e_0, e_1, e_2, e_3\}$. Observe that for all
      $0 \leq i \leq 3$, $\alpha_i + \alpha_{(i+1) \; \mod \; n} =
      \pi$ and $\intersect(e_i, \eplus) = \{ v \}$, hence
      $\coincide(e_0, e_2) \wedge \coincide(e_1, e_3)$.  In all loop
      iterations $\preRay = \postRay = \false$ thus only vertex $\{ v
      \}= P$ is generated (line 11).
    \item Consider $E = \{e_0, e_1, e_2\}$. Observe that for all $i$,
      $\ang{e_i}{\eplus} < \pi$, otherwise by Lemma \ref{lemRays},
      $\ray( \lbag E \rbag) \neq \emptyset$.  Thus in each iteration
      $i$, $\postRay = \preRay = \false$ and $P = \{ v \} =
      \intersect(e_i, \eplus)$ is generated.  Hence $\lbag E \rbag =
      \{ v \} + \cone(\{ \langle 0,0 \rangle \})$.
    \end{itemize}
  \item Suppose $E$ is one dimensional and $\dray(\lbag E \rbag) =
    \emptyset$.  Since the boundary of $\lbag E \rbag$ contains
    infinitely many points and $|E| = n$ (which is finite), there
    exists $e \in E$ such that $p_1, p_2 \in \lbag E \rbag$ where $p_1
    \neq p_2$ and $p_1$ and $p_2$ saturate $e$.  In fact there are
    infinitely many boundary points on the line between $p_1$ and
    $p_2$ which saturate $e$.  Observe $\lbag E \rbag$ contains no
    interior points, hence $\lbag E \rbag$ consists only of boundary
    points. Therefore there exists $e' \in E$ that saturates
    infinitely many of these points and for which $\coincide(e, e')$
    holds. As we assume that $\dray(\lbag E \rbag) = \emptyset$, $|E|
    < 2$.  Due to non-redundancy, there exists at most one $e_i \in
    E$ with $0<\ang{e}{e_i}<\pi$. Similarly for $e'$.  Hence $3 \leq
    |E| \leq 4$ follows.
    \begin{itemize}
    \item Consider $E = \{ e_0, e_1, e_2 \}$. Let $i \in [0,2]$ such
      that $\coincide(e_i, \eplus)$ holds. On iteration $i$ the loop
      generates a ray $r$ along the boundary of $\lbag e_i \rbag$ in
      line 10. A collinear ray is generated in iteration $(i+1) \;
      \mod \; n$ for $\eplus$ on line 9.  It is in this and iteration
      $(i+2) \; \mod \; n$ where $\{ v\} = \intersect(\eplus,
      \eplus[2]) = \intersect(e_{(i+2)\; \mod \; n}, e_i)$ is added to
      $P$. Thus $\lbag E \rbag = \{ v \} + \cone(\{ r \})$.
    \item Consider $E = \{ e_0, e_1, e_2, e_3 \}$. Let $i \in [0,3]$
      such that $\coincide(e_i, \eplus[2])$ holds.  In all four
      iterations $\preRay = \postRay = \false$ holds, resulting in two
      vertices resulting from the intersection of adjacent
      inequalities: $\{v_1\}\! =\! \intersect(e_i, \eplus) =
      \intersect(\eplus, \eplus[2])$, $\{v_2\}\! =\!
      \intersect(\eplus[2], \eplus[3]) = \intersect(\eplus[3], e_i)$.
      Hence $\lbag E \rbag = \conv(\{ v_1, v_2 \}) + \cone(\{ \langle
      0,0 \rangle \})$.
    \end{itemize}
  \item Suppose that $E$ is two dimensional and that none of the
    preceeding cases apply.
    \begin{itemize}
    \item Since $\coincide$ never holds for $|E| > 4$ the previous
      cases deal with all $e, e' \in E$ where $\coincide(e,e')$ holds.
      Because $\lbag E \rbag$ does not consist of a single point
      described by three inequalities, it must be two dimensional.
    \item Let $v \in V = \ext(\lbag E \rbag)$.  By Lemma
      \ref{lemInjection} there exists $e_i \in E$ such that
      $\ang{e_i}{\eplus} < \pi$ and $\{v\} = \intersect(e_i, \eplus)$.
      Then $\postRay$ is false in iteration $i$, hence $v \in P$, thus
      $V \subseteq P$. By Lemma \ref{lemIntersect} $P \subseteq \lbag
      E \rbag$.  If for all $i$,
      $\ang{e_i}{\eplus} < \pi$ holds then 
$\lbag E \rbag = \conv(P) + \cone(\langle 0,0 \rangle)$.
Otherwise let $i \in [0,m-1]$ such that $\ang{e_i}{\eplus} \geq \pi$. Then
 $\postRay$ will be true in
      iteration $i$ and $\preRay$ will hold in iteration $(i+1) \;
      \mod \; n$ yielding rays that are collinear to
$r_1 = \langle -b_i, a_i \rangle$ and $r_2 =
      \langle b_{(i+1) \; \mod \; n}, -a_{(i+1) \; \mod \; n}
      \rangle$.  Since the previous cases do not apply, whenever
      $\ang{e_i}{\eplus} = \pi$, $|E| \geq 3$ hence Lemma
      \ref{lemRays} can be applied. It follows that $\ray(\lbag E
      \rbag) = \cone(\{ r_1, r_2 \})$, hence $\lbag E \rbag = \conv(P)
      + \cone(\{r_1, r_2 \})$.
    \end{itemize}
  \end{itemize}
\end{proof}

\subsection{\label{sec-reconstruct}Reconstruction}
One advantage of the point and ray representation (over one that makes
lines explicit) is that $\extreme$ naturally generates lines as two
opposing rays in independent iterations, thus an explicit line case is
not required.  The remainder of the $\hull$ function combines the
points and rays of the two input polyhedra to construct a
corresponding set of inequalities.  The advantage of the simplified
representation carries over to the reconstruction phase in that
opposing rays from different polyhedra need not be recognised and
reconstituted as a line.

\begin{theorem}\label{theo-hull}
  Given $E_1, E_2 \subset \Lin$ be non-empty, finite, satisfiable and
  non-redundant, $\res = \hull(E_1, E_2)$ is non-redundant and the
  smallest $\res \subset \Lin$ with $\lbag \res \rbag \supseteq \lbag
  E_1 \rbag \cup \lbag E_2 \rbag$ such that for all $E \in \Lin$,
  $\lbag E \rbag \supseteq \lbag E_1 \rbag \cup \lbag E_2 \rbag
  \Rightarrow \lbag \res \rbag \subseteq \lbag E \rbag$.
\end{theorem}

\begin{proof} 
  The case for $E_1 = \emptyset$ or $E_2 = \emptyset$ on line 27 is
  trivial, thus assume $E_1 \cup E_2 \neq \emptyset$ and that lines
  28--35 are executed. Note that for all $r \in R$, $|r| = 1$ due to
  the normalisation in $\extreme$. Thus $(P + \{ 2\sqrt{2}sr \mid r
  \in R \}) \cap (-s,s)^2 = \emptyset$. Hence after line 12 the
  predicate $\inBox(s,p)$ holds whenever $p \in P = Q \setminus (P +
  \{ 2\sqrt{2}sr \mid r \in R\})$.  Line 38 handles the zero
  dimensional case, hence from line 39 onwards $|Q|>1$.  The
  arithmetic mean $q_p$ of all points is calculated in line 40. This
  point serves as reference when comparing two points for
  counter-clockwise ordering.  Observe that $q_p \in \conv(Q)$, in
  particular $q_p$ is not on its boundary if $\conv(Q)$ is two
  dimensional. The latter ensures for all boundary points $ q_1, q_2
  \in \conv(Q)$, $\theta(\overline{q_p, q_1}) \neq
  \theta(\overline{q_p, q_2})$, thus line 18 yields a total ordering
  on the boundary points in the two dimensional case.  Line 44
  performs the classic Graham scan which identifies the vertices of
  $\conv(Q)$.
  
  \begin{itemize}
  \item To show $\lbag \res \rbag \supseteq \lbag E_1 \rbag \cup \lbag
    E_2 \rbag$. In particular, show $\lbag \res \rbag \supseteq \lbag
    E_1 \rbag$, i.e.~for all $e \in \res$ , $\lbag \{ e \} \rbag
    \supseteq \lbag E_1 \rbag$.
    \begin{itemize}
    \item To show $\conv(Q) \subseteq \lbag \{ e \} \rbag$.  Suppose
      $e$ is added in line 60.  Then the flag $\add$ was true and $e =
      \con(q_{k_i}, q_{k_{(i+1) \; \mod \; m}})$.  For the sake of a
      contradiction, suppose there exists $q_{k_{j}} \notin \lbag \{ e
      \} \rbag$ such that $j \notin \{ i, {(i+1) \; \mod \; m} \}$.
      W.l.o.g.  let $\theta(\overline{q_p, q_{k_{j}}}) <
      \theta(\overline{q_p, q_{k_i}})$.  There exists a line through
      $q_p$ and $q_{k_{j}}$ which intersects the boundary of $e$ at a
      point $z$.  Then $q_{k_{j}} \in \conv(\{ z, q_{k_{(i+1) \; \mod
          \; m}} \})$ which contradicts that $q_{k_{i}}$ is a vertex.
      Since all $q_{k_0}, \ldots, q_{k_{m-1}} \in \lbag \{ e \} \rbag$
      it follows that $\conv(Q) \subseteq \lbag \{ e \} \rbag$.
      Furthermore it is easy to verify that the inequality $e$ added
      in line 58 or 59 holds for both $\{ q_{k_0}, q_{k_1} \} = Q$,
      hence $\conv(Q) \subseteq \lbag \{ e \} \rbag$.
    \item To show $\cone(R) \subseteq \ray(\lbag \{ e \} \rbag)$.
      \begin{itemize}
      \item Suppose $m>2$.  If either add was set at line 50 or
        line 54, there exists $q_j \in Q$ which saturates $e$ such
        that $\inBox(s, q_j)$ holds, hence $q_j \in P$.  Let $r \in
        R$.  Then $q_j + 2\sqrt{2}sr \in \conv(Q) \subseteq \lbag \{ e
        \} \rbag$.  Since $q_j$ saturates $e$, it follows that $r \in
        \ray(\lbag \{ e \} \rbag)$.
      \item Suppose $m=2$. Assume that $\inBox(s, q_{k_0})$ and
        $\inBox(s, q_{k_1})$ both do not hold, then there exist $p \in
        P$ and $r \in R$ with $q_{k_0} = p + 2 \sqrt{2} s r$. Note
        that since $m=2$, $p$ saturates $e$.  Continue as above.
        Assume $e$ was added in lines 58--59. Observe that $q_j=q_{k_i}$
        saturates $e$. Again, continue as in the first case.
      \end{itemize}
    \end{itemize}
    Thus $\lbag E_1 \rbag = \conv(P_1) + \cone(R_1) \subseteq \conv(P)
    + \cone(R) \subseteq \lbag \{ e \} \rbag$.  Similarly for $E_2$.
    Thus $\lbag \res \rbag \supseteq \lbag E_1 \rbag \cup \lbag E_2
    \rbag$.
    
  \item To show for all $E \subset \Lin$, $\lbag E \rbag \supseteq
    \lbag E_1 \rbag \cup \lbag E_2 \rbag \Rightarrow \lbag \res \rbag
    \subseteq \lbag E \rbag$. For the sake of a contradiction suppose
    there exists $p \in \lbag \res \rbag$ such that $p \notin \conv(P)
    + \cone(R)$. Hence for all $e \in \res$, $p \in \lbag \{ e \}
    \rbag$. Let $p' \in \conv(P) + \cone(R)$ such that $|p - p'|$ is
    minimal. Observe that $p'$ is unique due to convexity.

    \begin{itemize}
      
    \item Suppose $p' \in \ext(\conv(P) + \cone(R))$.  Hence there
      exists $i \in [0, m-1]$ such that $p' = q_{k_i}$.  Assume that
      $\inBox(s, q_{k_i})$ does not hold.  Then there exists $p'' \in
      P$, $r \in R$ and $\lambda >0$ such that $q_{k_i} = p'' +
      \lambda r$. Since $\lbag \res \rbag \supseteq \conv(P) +
      \cone(R)$, $p'' + 2 \lambda r \in \lbag \res \rbag$, thus
      $q_{k_i} \in \conv(\{p'', p'' + 2 \lambda r\})$ which is a
      contradiction, hence $\inBox(s,q_{k_i})$ holds.  Thus the flag
      $\add$ is set on line 54 in loop iteration $(i-1) \; \mod \; m$
      and $i$, hence the inequalities $e_{(i-1) \; \mod \; n} = \overline{q_{k_{(i-1)
            \; \mod \; m}}, q_{k_i}}$ and $e_i = \overline{q_{k_i},
        q_{k_{(i+1) \; \mod \; m}}}$ are added to $\res$.
      \begin{itemize}
      \item Suppose $\ang{e_{i-1}}{e_i}<\pi$. Then $\{ q_{k_i} \} =
      \intersect(e_{i-1}, e_i)$. Due to convexity $|q_{k_{(i-1) \;
      \mod \; m}} - p | > |q_{k_i} - p|$ and $|q_{k_{(i+1) \; \mod \;
      m}} - p | > |q_{k_i} - p|$. Hence $p' = q_{k_i}$ is the closest
      point to $p$ in the space $\lbag \{ e_{(i-1) \; \mod \; n}, e_i
      \} \rbag$. But this implies that $p \notin \lbag \res \rbag$
      which is a contradiction.

      \item Suppose $\ang{e_{i-1}}{e_i}=\pi$. Then $q_{k_{(i-1) \;
            \mod \; m}} = q_{k_{(i+1) \; \mod \; m}}$, thus $m=2$.
        Note that $p \in \lbag \{ e_{i-1}, e_i\} \rbag$ otherwise $p
        \notin \res$. Since $q_{k_i} \in P$, $\inBox(s, q_{k_i})$
        holds and line 58 or 59 adds an inequality $e \in \res$ with
        $\{ q_{k_i} \} = \intersect(e_{i-1}, e) = \intersect(e, e_i)$.
        Observe that $p \notin \lbag \{ e \} \rbag$ otherwise $p \in
        \conv(\{q_{k_i}, q_{k_{(i+1) \; \mod \; m}} \})$, thus $p
        \notin \lbag \res \rbag$.
      \end{itemize}
    \item Suppose $p' \notin \ext(\conv(P) + \cone(R))$, thus $p'$ is
      a boundary point of $\conv(P) + \cone(R)$. There exists a line
      $L \subset \real^2$ such that $p' \in L$ and $\conv(P) +\cone(R)
      \setminus L$ is convex (but not closed).  There exists a loop
      iteration $i$ such that the boundary of $e=\con(q_{k_i},
      q_{k_{(i+1) \; \mod \; m}})$ is exactly $L$.  Since $\conv(P) +
      \cone(R) \subseteq \lbag \{ e \} \rbag$, $p \notin \lbag \{ e \}
      \rbag$.
      \begin{itemize}
      \item Suppose $\inBox(s, q_{k_i})$ or $\inBox(s, q_{k_{(i+1) \;
            \mod \; m}})$ holds.  The flag add is true, thus $e \in
        \res$ and $p \notin \lbag \res \rbag$.
      \item Suppose neither $\inBox(s, q_{k_i})$ nor $ \inBox(s,
        q_{k_{(i+1) \; \mod \; m}})$ holds.  There exists $q_j \in P$
        such that $q_{k_i} = q_j + \lambda r$ for some $\lambda>0$ and
        $r \in R$.  Due to the ordering of the points, $k_i < j <
        k_{(i+1) \; \mod \; m}$ whenever $\conv(P) + \cone(R)$ is two
        dimensional.  Thus $\add$ is set in line 54.  If $\conv(P) +
        \cone(R)$ is one dimensional, $m=2$ and $\add$ is set in line
        49.  In both cases $e \in \res$ and thus $p \notin \lbag \{ e
        \} \rbag$.
      \end{itemize}
    \end{itemize}
  \end{itemize}
  
  It remains to show that $\res$ is non-redundant.  Note that the loop
  at lines 46--62 iterates once for each vertex $q_{k_i}$, creating
  inequalities $e_i = \overline{q_{k_i},q_{k_{(i+1) \; \mod \; m}}}$.
  For all $i \neq j$, $\theta(e_i) \neq \theta(e_j)$ due to the fact
  that $q_{k_0}, \ldots, q_{k_{m-1}}$ are consecutive vertices of the
  hull of $\conv(Q)$, thus $\{e_0, \ldots, e_{m-1}\}$ has no
  redundancies.  Now consider the inequality $e_i'$ added at line 58
  or 59. Since $m=2$, $\ang{e_i}{e_{(i+1) \; \mod \; m}} = \pi$ and
  since $\theta(e_i) < \theta(e_i') < \theta(e_{(i+1) \; \mod \; m})$,
  the inequality $e_i'$ is not redundant. Hence $\res \subseteq \{e_0,
  e_0', \ldots, e_{m-1}, e_{m-1}' \}$ is non-redundant.
\end{proof}

The running time of the algorithm is $O(n \log n)$ where $n= |E_1| +
|E_2|$. After the sorting step at line 2 in $\extreme$, each
inequality generates at most two rays and one point. Thus $\extreme$
is in $O(n \log n)$.  The flag $\preRay$ is true if the angle between
two consecutive inequalities is at least $\pi$. Thus $\preRay$ can
only be true in at most two loop iterations. Similarly for $\postRay$.
Hence $\extreme$ returns at most four rays for each polyhedron, thus
$|R|\leq 8$ at line 31 and $O(|Q|) = O(n)$. The dominating cost in the
$\hull$ function is the sorting step at line 42 which we assume is in
$O(n \log n)$. The scan is linear \cite{graham72efficient} and
partitions the point set $Q$ into vertices and non-vertices. The loop
at lines 46--62 runs once for each vertex, whereas the inner loop at lines
52--56 runs at most once for each non-vertex. It follows that the overall
running time is in $O(n \log n)$.

\section{\label{sec-conclusion}Conclusion}

An $O(n \log n)$ algorithm for calculating the convex hull of planar
$H$-polyhedra has been presented, thereby improving on existing
approaches.  The algorithm applies a novel box construction which reduces
the problem to calculating the convex hull of a set of points.
Implementing the algorithm exposed a number of subtleties which
motivated a complete proof of the algorithm.




\begin{thebibliography}{10}

\bibitem{anderson87reevaluation}
Kenneth~R. Anderson.
\newblock A {R}eevaluation of an {E}fficient {A}lgorithm for {D}etermining the
  {C}onvex {H}ull of a {F}inite {P}lanar {S}et.
\newblock {\em Information Processing Letters}, 7(1):53--55, January 1978.

\bibitem{avis92pivoting}
D.~Avis and K.~Fukuda.
\newblock A {P}ivoting {A}lgorithm for {C}onvex {H}ulls and {V}ertex
  {E}numeration of {A}rrangements and {P}olyhedra.
\newblock {\em Discrete Computational Geometry}, 8:295--313, 1992.

\bibitem{chernikova68algorithm}
N.~V. Chernikova.
\newblock Algorithm for {D}iscovering the {S}et of {A}ll {S}olutions of a
  {L}inear {P}rogramming {P}roblem.
\newblock {\em {USSR} {C}omputational {M}athematics and {M}athematical
  {P}hysics}, 8(6):282--293, 1968.

\bibitem{graham72efficient}
R.~L. Graham.
\newblock An {E}fficient {A}lgorithm for {D}etermining the {C}onvex {H}ull of a
  {F}inite {P}lanar {S}et.
\newblock {\em Information Processing Letters}, 1(4):132--133, 1972.

\bibitem{klee59some}
V.~L. Klee.
\newblock Some characterizations of convex polyhedra.
\newblock {\em Acta Mathematica}, 102:79--107, 1959.

\bibitem{verge92note}
H.~{Le Verge}.
\newblock A {N}ote on {C}hernikova's algorithm.
\newblock Technical Report 1662, Institut de Recherche en Informatique, Campus
  Universitaire de Beaulieu, France, 1992.

\bibitem{motzkin53double}
T.~S. Motzkin, H.~Raiffa, G.~L. Thompson, and R.~M. Thrall.
\newblock The {D}ouble {D}escription {M}ethod.
\newblock In {\em Contributions to the Theory of Games}, number~28 in Annals of
  Mathematics Study. Princeton University Press, 1953.

\bibitem{preparata85computational}
F.~P. Preparata and M.~I. Shamos.
\newblock {\em Computational Geometry}.
\newblock {T}exts and {M}onographs in {C}omputer {S}cience. {S}pringer
  {V}erlag, 1985.

\bibitem{sedgewick98algorithms}
R.~Sedgewick.
\newblock {\em Algorithms}.
\newblock Addison-Wesley, 1988.

\bibitem{seidel97handbook}
R.~Seidel.
\newblock Convex {H}ull {C}omputations.
\newblock In J.~E. Goodman and J.~O'Rourke, editors, {\em Handbook of Discrete
  and Computational Geometry}, pages 361--376. CRC Press, 1997.

\bibitem{ziegler}
G.~Ziegler.
\newblock {\em Lectures on Polytopes}, volume 152 of {\em Graduate Texts in
  Mathematics}.
\newblock Springer-Verlag, 1994.

\end{thebibliography}
\end{document}